\documentclass[twocolumn]{aastex631}
\usepackage{fontawesome}

\begin{document}

\title{Nuggets of Wisdom: Determining an Upper Limit on the Number Density of Chickens in the Universe}

\correspondingauthor{Rachel Losacco}
\twitter{RachelDoesAstro}

\author{Rachel Losacco}
\affiliation{University of Florida, 211 Bryant Space Science Center, Gainesville, FL 32611}

\author{Zachary Claytor}
\affiliation{University of Florida, 211 Bryant Space Science Center, Gainesville, FL 32611}

\begin{abstract}

The lower limit on the chicken density function (CDF) of the observable Universe was recently determined to be approximately 10$^{-21}$ chickens pc$^{-3}$. For over a year, however, the scientific community has struggled to determine the upper limit to the CDF. Here we aim to determine a reasonable upper limit to the CDF using multiple observational constraints. We take a holistic approach to considering the effects of a high CDF in various domains, including the Solar System, interstellar medium, and effects on the cosmic microwave background. We find the most restrictive upper limit from the domains considered to be 10$^{23}$ pc$^{-3}$, which ruffles the feathers of long-standing astrophysics theory.

\end{abstract}

\date{April 1, 2023}

\section{Introduction} \label{sec:intro}

The chicken density function (CDF) entered the scientific spotlight in March 2022 when a listener of the podcast \textit{Dear Hank \& John} wrote in with the question: 
``Do we have any proof that the space between galaxies isn't just filled with a bunch of chickens?" 
Host Hank Green and Roman Mars \citep{Green2022} conjecture that the upper limit would be constrained by the distance at which the chickens could see each other, though it could be more than two chickens per cubic light year, or about 70 pc$^{-3}$.
Finally, they formulate what we believe should be a leading scientific question in the next decadal survey: 
``There is a number of chickens that could be in the intergalactic medium that we wouldn't notice... How many chickens would it have to be before we notice?"

Reddit user u/TheStig465 goes on to determine the lower limit of the CDF to be $2.13\times10^{-21}$pc$^{-3}$, or $6.15\times10^{-23}$ chickens ly$^{-3}$, given the volume of the observable Universe to be $4.21\times 10^{32}$ ly$^3$ and the Earth's chicken population to be $25.9\times10^9$ chickens \footnote{\url{https://www.reddit.com/r/theydidthemath/comments/tvqqqh/self_how_many_chickens_exist_per_cubic_lightyear/}}. 
This lower limit, by definition, assumes the only chicken in the Universe are those on Earth.

Recent work\footnote{\url{https://isotropic.org/papers/chicken.pdf}} has highlighted how prominent the species is to our fundamental understanding of the Universe. Although observations of \textit{Homo sapien sapien} in space have been recorded as early as 1961 (e.g., Gagarin 1961), \textit{Gallus gallus domesticus} outnumbers this species by a factor of four on Earth; this abundance ratio may remain constant on an intergalactic scale. 

This line of thinking leads to how substantial of an effect chickens would have on multiple scales. From Mercury's orbit to the asteroid belt to the Oort cloud, the interaction between bodies of the Solar System and undetectable chickens can explain phenomena that otherwise rely on the superfluous theory of general relativity. A more accurate estimate of the CDF can dramatically alter photometric extinction curves and measurements of ISM metallicity. The answer to this question can also reshape astronomical standard candles as we know them, potentially even resolving standing crises like the Hubble tension.

In this paper, we explore the upper limit of the CDF using three approaches: as asteroid-like objects within our Solar System (Section \ref{sec:solarsystem}), components of the interstellar medium and intergalactic medium via photometric extinction (Section \ref{sec:extinction}), and on a cosmic scale as it may affect background signal (Section \ref{sec:CMB}). These factors and more are considered in Section \ref{sec:results} in order to determine a resulting upper limit. Section \ref{sec:conclusion} provides a concluding thoughts and reflection on future work.

\section{Methods}

We focus on three realms of chicken observations: the effects of individual chickens in the Solar System, the impacts of a chicken-based interstellar medium, and the cosmic background radiation of a high CDF.

\subsection{Solar System Chickens}\label{sec:solarsystem}

\textit{Why did the chicken cross the asteroid belt?}
Answer: solar radiation pressure and the Yarkovsky effect \citep[e.g.,][]{Bottke2006}. These phenomena describe the interaction of highly reflective asteroids in the asteroid belt and solar radiation. Solar radiation pressure slowly perturbs asteroid orbits outward, but the Yarkovsky effect is more subtle. Rotating asteroids absorb some sunlight on the sunward side. As they rotate, they re-radiate in a different direction, which transfers some momentum from the asteroid. If the asteroid re-radiates in the direction of motion, it slows the object, perturbing its orbit outward. The combined effects from radiation pressure and Yarkovsky could perturb asteroids into a region which is in resonance with Jupiter, disrupting their orbits dramatically and ultimately dislodging them from the asteroid belt altogether. These asteroids can then become meteors, colliding with other objects in the Solar System such as the Moon and Earth.

Chickens within the Solar System, especially those in and around the asteroid belt, would experience similar effects. The reflectance of the plumage of white Oakham Blue hens ranges from 80\% to $>$90\% for their wing, tail, rump, back, and neck in the visible spectral range \citep{Bright2007}. This results in a comparable albedo to asteroids affected by solar radiation pressure and the Yarkovksy effect. Therefore, if a high abundance of chickens populate the asteroid belt, one would expect to see chicken meteors and their impacts throughout the Solar System. At time of writing, there has been no recorded evidence of such meteors, concluding that the asteroid belt is not well populated with high-albedo chickens.

A final consideration is low-albedo chickens. \citet{Bright2007} also measured the reflectance of the grey and black varieties of Oakham Blue hens. The former exhibit 50\%-60\% reflectance in the visible spectrum, while the latter reach as low as 10\% reflectance. These ``dark chickens"\footnote{Not to be confused with the dark meat of a chicken} could therefore remain in the asteroid belt undetected and unaffected by the phenomena described above.

Dark chickens may also be present in the inner Solar System, flying under the radar of current observational technology. In Section \ref{sec:extinction}, Equation \ref{equ:flux} concludes that up to $2\times10^{18}$ chickens AU$^{-3}$ may be present within the 0.01\% precision of detection capabilities \citep{Kopp2011}. While assumed to be uniform, perturbations in the distribution in and around Mercury's orbit may be able to account for its precession, which was otherwise attributed to the superfluous theory of general relativity \citep{will_1993}. Further implications of dark chickens are considered in Section \ref{sec:conclusion}. While astronomers are encouraged to continue exploring Mercury's orbit for evidence of dark chickens, it is strongly recommended for the US Department of Defence and Space Force to thoroughly examine the possibilities of dark chickens occupying low- and high-Earth orbits.


\subsection{Detection by Photometric Extinction}\label{sec:extinction}
Here we consider the detection of interstellar chickens via extinction. We assume spherical chickens with average radius $a'$. In the simple (nonrealistic) case of non-overlapping occulting chickens, the flux $\delta f$ extinguished from a source, expressed as a fraction of the source's total flux, is
\begin{equation} 
    \delta f = \left(\frac{a'}{R}\right)^2 \sum_{i=1}^{N} \left(\frac{z_i'}{d}\right)^{-2},
\end{equation}
where $R$ is the radius of the source, $N$ is the total number of chickens occulting the source, $z_i'$ is the distance to chicken $i$, and $d$ is the distance to the source. For simplicity we scale the radius and distance to each chicken, defining $a = a'/R$ and $z_i = z_i'/d$, yielding
\begin{equation}
    \delta f = a^2 \sum_{i=1}^{N} \frac{1}{z_i^2}.
\end{equation}
To avoid the regime of single-object occultation, we assume $\frac{a'}{z'} \ll \frac{R}{d}$, or equivalently $a \ll z$. For sufficiently large $N$, the sum approaches $N$ times the expected value of the inverse square distance, and the extinguished flux becomes
\begin{equation} \label{flux}
    \delta f = a^2 N \left\langle\frac{1}{z^2}\right\rangle.
\end{equation}
We must now evaluate $N$, the total number of chickens occulting the source; and the expected value of $1/z^2$, which depends on the distribution of $z_i$, the distance to each chicken.

\begin{figure*}
    \centering
    \includegraphics[width=0.7\textwidth]{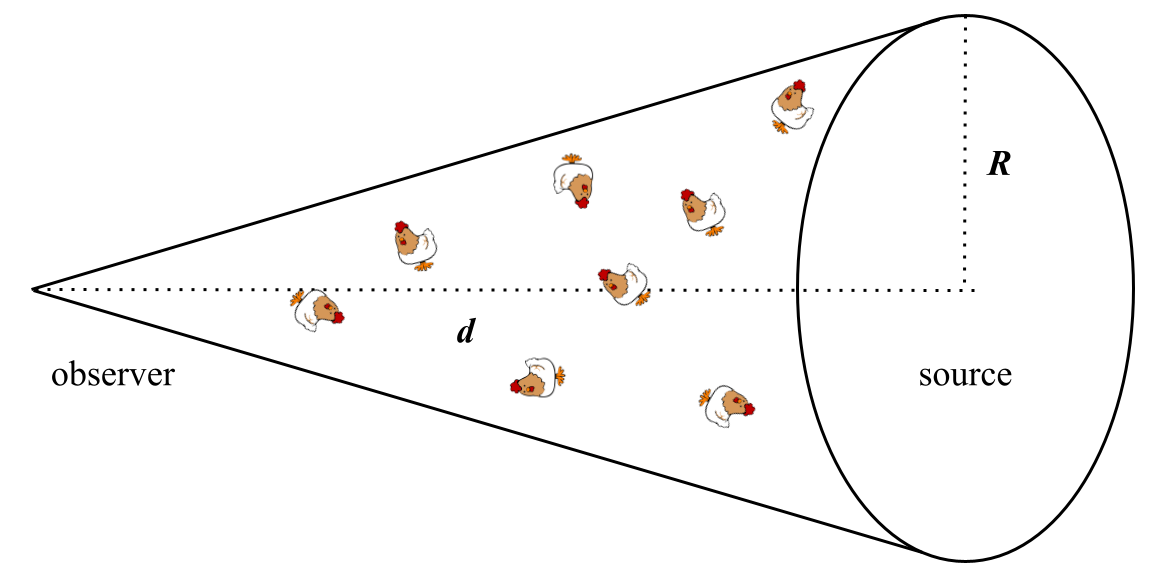}
    \caption{Schematic of the observed volume, which can be represented by a cone. Here $R$ is the source radius, and $d$ is the distance to the source.}
    \label{fig:frustum}
\end{figure*}

Figure~\ref{fig:frustum} shows a schematic diagram of the volume between the source and the observer, which can be represented by a cone with end radius $R$ and length $d$. The volume of such a cone is $\frac{1}{3} \pi d R^2$. Assuming a uniform spatial number density $n$ of chickens, the number $N$ of chickens occulting the source is then

\begin{equation}
    N = \frac{1}{3} \pi d R^2 n.
\end{equation}
We emphasize that $n$ is the number density we want to constrain.

The expected value of $1/z^2$ over $a < z < b$ is given by
\begin{equation}
    \left\langle\frac{1}{z^2}\right\rangle = \int_a^b \frac{1}{z^2} p(z) \mathrm{d}z,
\end{equation}
where $p(z)$ is the probability density function of the distance $z$. Since chickens are assumed to be distributed uniformly across the conical observation volume, $p(z)$ must scale with the area $A$ of the conic cross section, given by $A = \pi r^2$ with $r/z = R$ (recall that $z = z'/d$ and $z$ is unitless). Therefore, $p(z) \propto z^2$. Normalizing over the domain $0 < z < 1$ yields $p(z) = 3z^2$. The expected value of 1/$z^2$ is then
\begin{equation}
    \left\langle\frac{1}{z^2}\right\rangle = \int_0^1 3 \mathrm{d}z = 3,
\end{equation}
corresponding to a distance of $z = 3^{-1/2} \approx 0.58$. Redefining $a' \rightarrow a$ such that $a' = a/R$, Equation~(\ref{flux}) then becomes
\begin{equation}\label{equ:flux}
    \delta f(n) = \pi d (a')^2 R^2 n = \pi d a^2 n,
\end{equation}
where now $a$ is the average radius of a chicken, $d$ is the distance to the source, and $n$ is the CDF, the spatial number density of chickens. Whether by convenience or by divine providence, the average radius of a chicken is approximately $\pi^{-1/2}$ m, so this simplifies further to
\begin{equation}
    \delta f(n) = (1~\mathrm{m}^2) n d.
\end{equation}
For the Sun, for which we can measure the total solar irradiance to about 0.01\% precision \citep{Kopp2011}, this yields an upper limit of $n \leq 2\times10^{18}~\mathrm{AU}^{-3}$ or $2\times10^{34}~\mathrm{pc}^{-3}$. 

More distant objects provide stronger constraints on the CDF provided photometric precision does not decrease faster than the distance increases. For example, the brightness of stars at the tip of the red giant branch (TRGB) can be measured with 0.05\% precision using the Hubble Space Telescope \citep{Anand2021}. At extreme distances on Mpc scales, the CDF must be less than $n \leq 10^{23}~\mathrm{pc}^{-3}$ to go unnoticed by TRGB measurements. 

\subsection{Detection of the Chicken Meat Background}\label{sec:CMB}
We predict a measurable thermal Chicken Meat Background (CMB) for a sufficiently high CDF. A single chicken with temperature $T$ and distance $z$ would have a luminous flux of
\begin{equation}
    f = L/z^2 = \pi \sigma a^2 T^4 / z^2,
\end{equation}
where $\sigma$ is the Stefan-Boltzmann constant, and $a$ is again the average chicken radius. Again assuming non-overlapping chickens and no cosmological redshift dependence, the total flux from all chickens in the sky is
\begin{equation}
    F = \pi \sigma a^2 T^4 \sum_{i=1}^N z_i^{-2} = \pi \sigma a^2 T^4 N \left\langle\frac{1}{z^2}\right\rangle,
\end{equation}
applying the same large $N$ approximation as before, except now $z$ has absolute units of distance. Of course, in this approximation we run into Olbers' paradox, since $\left\langle z^{-2}\right\rangle$ is formally unbounded. We resolve this by taking into account the finite time in which chickens have existed on Earth. As we know them today, chickens were domesticated around 7,000--10,000 years ago \citep{Laatsch}. While chickens are fast, they have not been observed to travel faster than light, so this places a limit on the radius in which chickens could appear. We adopt a distance of 10,000 ly, bounding the expected value of $z^{-2}$. The CMB flux is then
\begin{equation}
     F = \pi \sigma a^2 T^4 N \int_0^b \frac{1}{z^2}p(z)\mathrm{d}z,
\end{equation}
where $b$ is the chicken radius limit of 10,000 ly. Now in a spherical volume, $N = \frac{4}{3}\pi b^3 n$, and $p(z) = 3z^2/b^3$, so this becomes
\begin{equation}
     F = 4\pi^2 \sigma a^2 T^4 b n.
\end{equation}
Note that the temperature of the chickens is important; whether the chickens are alive (i.e., $T = 300$ K) or dead (local equilibrium temperature, mostly 3 K) makes a substantial difference. For living chickens at $T = 300$ K within a radius $b = 10,000$ ly, the flux density across the entire sky would be
\begin{equation}
    f(n) = (850~\mathrm{erg~ly~s}^{-1}~\mathrm{as}^{-2}) n
\end{equation}
Realistically, the chickens would be much cooler in the vacuum of space, closer to the background temperature of 3 K, which reduces the value of $f$ by a factor of 10$^8$. In fact, if we suppose the cosmic microwave background is from cold, thermally glowing chickens, we can estimate the number density from the microwave background flux, which has a density of about 10$^{-3}$~erg~s$^{-1}$~cm$^{-2}$~sr$^{-1}$. This provides an upper limit on the CDF of $n \leq 10^{29}$ pc$^{-3}$.

\section{Results and Discussion}\label{sec:results}

\begin{table}
    \centering
    \begin{tabular}{c|c}
        \hline
        Constraint & $n_\mathrm{max}$ (pc$^{-3}$) \\
        \hline
        Solar System impacts & undetermined \\
        Solar extinction & 10$^{34}$ \\
        TRGB extinction & 10$^{23}$ \\
        CMB (Chicken Meat Background) & 10$^{29}$ \\
        \hline
        Adopted Upper Limit & 10$^{23}$ \\
        \hline
    \end{tabular}
    \caption{Estimated upper limits of the Chicken Density Function (CDF) from various regimes. We adopt the most strict limit as the likely upper limit to the CDF.}
    \label{tab:limits}
\end{table}

We summarize the CDF upper limit estimates in Table~\ref{tab:limits}. Adopting the strictest limit, we find that the CDF must be less than about $10^{23}$ pc$^{-3}$ (10$^{7}$ AU$^{-3}$). Higher than this, we would notice irregularities in the position of the tip of the red giant branch (TRGB) in distant galaxies. On the other hand, densities higher than this would cause appreciable photometric extinction in the location of the TRGB for which models do not account, affecting distance measurements. This might therefore give rise to the notorious tension in Hubble constant estimates between local- and early-Universe investigations \citep[e.g.,][]{Riess2022}.

We note that the upper limit of 10$^{7}$ AU$^{-3}$ underpredicts the number of chickens observed on Earth \citep[about 30 billion,][]{howmany}, implying Earth's population represents a large overdensity in the overall distribution of chickens. While we have assumed a homogeneous distribution of chickens, inhomogeneity at densities this high can have cosmic consequences. A region of overdensity of chickens may lead to gravitational collapse, exceeding the Jeans limit and creating a chicken star. Due to the high carbon and oxygen abundances, we expect that such a chicken star might observationally resemble a standard white dwarf, and we urge the type Ia supernova community to give serious consideration to chicken stars in addition to single- and double-degenerate scenarios. 

\citet{Jayasena2013} recently investigated why many foods taste like chicken. They found that the flavor comes mainly from a specific polycyclic aromatic hydrocarbon (PAH) primarily found in chicken: 2-Methyl-3-furanthiol. Further observations of interstellar PAHs are needed to measure the abundance of 2-Methyl-3-furanthiol, which is likely to be a tracer for the CDF. Additionally, sulfur-rich PAHs may be a strong indicator for early gravitational globules (EGGs), the protostellar stage of a chicken star's life cycle.

Chickens located in the habitable zone of planetary systems are likely to have an effective equilibrium temperature of 313K, and observations in these systems should consider this as the effective temperature for corresponding blackbody radiation. Theory also suggests a region around a star where the internal temperature of the chicken reaches 347K (165$^\circ$F), the temperature at which chicken is fully cooked and safe to eat. Much like the habitable zone is a region where liquid water can be found for human consumption, this region, known as the Kepler 165-Fahrenheit Convection (KFC) zone is where one can search for chicken that is safe for human consumption.

The rate at which chickens form, or the chicken formation rate (CFR), is determined by the fuel source and the CDF. The lower limit of the CFR is defined by a minimum interaction rate of chicken, while the upper limit is set by the Jeans limit. The introduction of \textit{Homo sapien sapien}, however, acts as a catalyst for both exponential production \citep{checkin} while also expediting the chicken's natural life cycle. Therefore, the presence of other species can greatly impact the chicken evolution, and should be taken into account when analyzing observations.

\pagebreak
\section{Conclusion}\label{sec:conclusion}
In this work we have constrained the upper limit on the Chicken Density Function (CDF), the number density of unobserved chickens in the observable Universe. We have followed Solar System, interstellar, intergalactic, and cosmological considerations. We take the most restrictive of these limits to be the current best upper limit: 10$^{23}$ chickens per cubic parsec (10 million per cubic AU), constrained by the photometric precision of tip-of-red-giant-branch stars in faraway galaxies. 

While we have considered a plethora of scenarios across a vast range of cosmic distances, there are several scenarios we have not considered which may further constrain the upper limit to the CDF. For example, particularly low albedo chickens could avoid detection while contributing to the mass of gravitationally bound systems, acting as what we might call dark matter. We therefore propose two new modes of dark matter: Weakly Interacting Nuggets of Gravity (WINGs) and Celestial Hydrodynamically Interacting Chickens (CHICs). Another consideration is whether such low-albedo chickens could coalesce into a black hole. Such chicken black holes may exert pressure on cosmic scales, giving rise to dark-energy-like phenomena \citep{Farrah2023}. Further constraints on the CDF will require new observations, new podcast episodes, and for new theories to be hatched.

\begin{acknowledgments}
We thank John and Hank Green, Roman Mars, and Reddit user u/TheStig465 for their insight, as well as Gagandeep Anand for useful discussions that improved the quality of this paper.
\end{acknowledgments}

%






\bibliography{sample631}{}
\bibliographystyle{aasjournal}



\end{document}